\documentclass[preprint, 5p]{elsarticle}

\usepackage[english]{babel}

\usepackage{amsmath}
\usepackage{amssymb}
\usepackage{booktabs}
\usepackage{graphicx}

\usepackage{siunitx}
\DeclareSIUnit\atomicmassunit{u}
\DeclareSIUnit{\count}{counts}
\DeclareSIUnit{\primary}{\text{primary particles}}
\usepackage{comment}
\usepackage{subcaption}
\usepackage[colorlinks=true, allcolors=blue]{hyperref}
\usepackage{booktabs,tabularx,siunitx}
\newcolumntype{L}{>{\raggedright\arraybackslash}X}
\usepackage{tikz}
\usepackage{float}

\newcommand{\replyTo}[1]{\textit{** Comment #1 ** }}
\renewcommand{\replyTo}[1]{}

\journal{Life Sciences in Space Research}

\begin{document}

\begin{frontmatter}

\title{Hybrid Active-Passive Galactic Cosmic Ray Simulator:\\in-silico design and optimization}

\author[1,2]{L.~Lunati}
\ead{L.Lunati@gsi.de}
\author[1]{E.~Pierobon}
\ead{E.Pierobon@gsi.de}
\author[1,3]{U.~Weber}
\ead{u.weber@gsi.de}
\author[1]{T.~Wagner}
\ead{T.Wagner@gsi.de}
\author[1]{T.~Pfuhl}
\ead{tabea.pfuhl@gmx.de}
\author[1,2,4]{M.~Durante}
\ead{M.Durante@gsi.de}
\author[1]{C.~Schuy\texorpdfstring{\corref{cor1}}{}}
\ead{C.Schuy@gsi.de}

\cortext[cor1]{Corresponding author}

\affiliation[1]{organization={Biophysics Department, GSI Helmholtzzentrum für Schwerionenforschung},
    addressline={Planckstraße 1},
    postcode={64291},
    city={Darmstadt},
    country={Germany}}
\affiliation[2]{organization={Institute of Condensed Matter Physics, Technische Universität Darmstadt},
    addressline={Hochschulstr. 6},
    postcode={64289},
    city={Darmstadt},
    country={Germany}}
\affiliation[3]{organization={Technische Hochschule Mittelhessen, University of Applied Sciences, Fachbereich LSE},
    addressline={Gutfleischstr. 3},
    postcode={35390},
    city={Gießen},
    country={Germany}}
\affiliation[4]{organization={Department of Physics ``Ettore Pancini'', University Federico II}, 
    addressline={Via Cintia, 21 - Building 6},
    postcode={80126},
    city={Naples},
    country={Italy}}

\begin{abstract}
\replyTo{6}High-energy heavy-ion particle accelerators have long served as proxies for the harsh space radiation environment, enabling both fundamental life-science research and applied testing of flight hardware. Traditionally, monoenergetic high-energy heavy-ion beams have been employed for practicality, providing valuable datasets that underpin radiation risk and predictive computational models. However, such beams cannot fully reproduce the mixed-field nature of space radiation, motivating the development of realistic analogs for improved risk assessment and countermeasure evaluation in preparation for future deep-space missions to Moon or Mars.
Spearheaded by developments at the NASA Space Radiation Laboratory, the GSI Helmholtzzentrum f{\"{u}}r Schwerionenforschung, supported by the European Space Agency (ESA), has established advanced space radiation simulation capabilities in Europe. Here, we present the design, optimization, and in-silico benchmarking of GSI's hybrid active-passive Galactic Cosmic Ray (GCR) simulator, together with a computationally optimized phase-space particle source for Geant4, which is available to external users for their own simulation studies and experimental planning. 
\end{abstract}


\end{frontmatter}

\section{Introduction}
\label{sec:Introduction}
\replyTo{1, 7, 8, 9}Space Radiation is one of the major obstacles to human exploration of the Solar System \cite{Chancellor2014}. With the renewed interest from space agencies worldwide in returning to the Moon within this decade, and with plans for Mars in the near future, space explorers and mission-critical electronic systems will face higher levels of radiation than those experienced in Low Earth Orbit (LEO). Outside the Earth’s protective magnetosphere, the radiation environment is dominated by Galactic Cosmic Rays (GCRs) and Solar Energetic Particles (SEPs) \cite{Durante2011}. Even though space habitats are engineered to ensure crew safety, chronic exposure to Galactic Cosmic Radiation (GCR) remains the most significant long-term health risk including carcinogenesis, degenerative tissue effects, and acute and late Central Nervous System (CNS) disorders \cite{Sishc2022}.
Understanding and mitigating these risks is essential to enable a safe and sustainable human presence in space, and is only possible after characterizing and reducing current biological and physical uncertainties associated with prolonged exposure to the complex mixed field of space radiation \cite{Durante2014, Fogtman2023}. 
Regardless of the specific space radiation source or mission scenario of interest, high-energy heavy-ion accelerators are indispensable tools for studying radiation effects and developing effective mitigation strategies. Traditionally, ground-based experiments have employed independent irradiations with monoenergetic single-ion beams. While such experiments provide only an incomplete analog of the complex space radiation environment, collectively they have yielded extensive datasets that underpin radiation risk assessment and the development of predictive computational models. However, this approach cannot fully capture the mixed nature of space radiation, where interactions from particles of different charge and energy occur in spatial and temporal proximity and may influence biological outcomes \cite{Raber2019}.
Advanced concepts for replicating SEPs \cite{LaTessa2016} and GCRs \cite{kim2015issues, chancellor2017targeted, chancellor2018limitations, timoshenko2017particle, gordeev2021new, gordeev2024computer} have been investigated, with the approach implemented at the National Aeronautics and Space Administration (NASA) Space Radiation Laboratory (NSRL) at Brookhaven National Laboratory (BNL), USA, being the most advanced operational system to date \cite{Simonsen2020}. NASA's “GCR simulator” employs fast ($\lesssim 2$ minutes) switching between 33 ion beams at different energies to generate a single fixed reference field approximating the GCR experienced behind \qty{20}{\gram\per\cm\squared} of aluminum shielding, under solar minimum conditions, in human blood-forming organs (BFO).\\
To enable realistic ground-based studies of space radiation exposure effects in Europe, the GSI Helmholtzzentrum f{\"{u}}r Schwerionenforschung (GSI), supported by the European Space Agency (ESA), developed irradiation tools that closely mimic both Solar Particle Events (SPEs) \cite{Pfuhl2024} and GCRs \cite{Schuy2020} adopting design strategies distinct from NASA and NSRL. Following the development of 3D range modulators for particle therapy \cite{Simeonov2017}, the GSI's SPE simulator uses a complex modulator interacting with a 220 MeV primary proton beam to mimic a SPE spectrum. The GSI's GCR simulator, instead, employs a hybrid active-passive approach, based on active switching of the energy of a $^{56}$Fe beam, combined with passive solid-slab modulators and highly structured, periodic, complex modulators. This enables the generation of a mixed radiation field that reproduces the GCR at 1 astronomical unit (\qty{}{\astronomicalunit}) in a lightly shielded habitat (\qty{10}{\gram\per\cm\squared} aluminum equivalent thickness), during 2010 solar minimum conditions.\\ 
Rather than defining a single fixed reference field, the GSI concept provides a flexible framework capable of reproducing a broad range of space-radiation environments. Starting from a common baseline field corresponding to the GCR-like spectra behind \qty{10}{\gram\per\cm\squared} Al, the irradiation environment can be further adapted by inserting additional materials of various compositions and thicknesses along the beamline. Different heliospheric scenarios can likewise be reproduced without any hardware modifications by appropriately re-weighting modulator configurations. This flexibility also enables replication of the NSRL baseline environment by placing appropriate shielding and tissue-equivalent phantom materials to simulate the self-shielding at BFO depth.\\ 
An extensive dataset of pre-simulated base and target data, along with a custom analytical optimizer, was used to determine material weights for possible modulator geometries. 
This paper focuses on the design and in-silico optimization of GSI's GCR simulator, intended as a realistic analog of the GCR environment. The simulator aims to reproduce the key features of space radiation within the practical and operational constraints of ground-based setups. Its implementation, together with the first experimental validation measurements, is described and discussed in \cite{pierobon2025exp}.

\section{Material and methods}
\label{sec:Material_and_methods}

\subsection{Galactic Cosmic Radiation Environment}
\label{ssec:Galactic_Cosmic_Radiation_Environment}

Galactic Cosmic Rays form an isotropic background of highly penetrating radiation originating outside the Solar System, likely from astrophysical explosive events such as supernovae, neutrons stars, pulsars, or other high energy phenomena. They consist of nuclei of the naturally occurring chemical elements, from hydrogen to uranium. In free space, about 98$\%$ of GCRs are protons and heavier ions, and about 2$\%$ are positrons and electrons; the latter contribute negligibly to space radiation exposure.  
Protons make up roughly 88$\%$ of the total flux, helium isotopes (mainly $^4$He) account for approximately 10$\%$, and high charge and energy (HZE) particles comprise the remaining 1--2$\%$ (abundances vary slightly depending on the phase of the solar cycle) \cite{townsend2020space}. The abundance of elements with $Z > 26$ decreases sharply and poses little health risk, making nuclei up to nickel ($Z=28$) the primary concern.
Although HZE particles represent only a small fraction of the total flux, they account for approximately 89$\%$ of the dose equivalent (Sv) in free space behind thin shielding (\qty{5}{\gram\per\cm\squared} Al). Iron, despite being only about one-tenth as abundant as carbon or oxygen, is the single largest contributor, responsible for approximately 26$\%$ of the total dose equivalent, due to its high charge and the correspondingly large quality factor \cite{Durante2011}. \\
GCRs span a broad energy range, with \qty{10}{\MeV\per\atomicmassunit} to \qty{10}{\GeV\per\atomicmassunit} being most relevant for radiobiological research and space mission planning. Within the solar system, the energy spectra of GCR are modulated by the solar wind, with their intensities being anti-correlated with solar activity. 
Solar modulation can cause GCR ion fluxes during solar maximum to be a factor 3-4 lower than during solar minimum, while exposure estimates behind a given amount of shielding are reduced by roughly a factor of 2 \cite{Townsend1990}.
As GCRs traverse spacecraft material, electromagnetic and nuclear interactions alter the deep space radiation field, originating a non-uniform in-habitat environment composed of attenuated primary and secondary particles, including energetic neutrons, protons, helium ions, and heavier ($Z \geq 3$) fragments \cite{Zeitlin2016}. 
To enable the optimization process of the GCR simulator, a standardized Monte Carlo simulation was performed using the Geant4 Monte Carlo toolkit \cite{Agostinelli2003, Allison2006, Allison2016} to generate reference spectra as optimization targets for the following steps. The free-space GCR environment at \qty{1}{\astronomicalunit}, as defined by the ESA-DLR GCR model \cite{Matthia2013}, was simulated for protons and heavy ions up to nickel ($Z=28$). The modeled kinetic-energy range spanned from \qty{10}{\MeV\per\atomicmassunit} to \qty{100}{\GeV\per\atomicmassunit}, assuming 2010 solar minimum conditions\replyTo{12}. Primary particles with energies below \qty{10}{\MeV\per\atomicmassunit} were excluded since their contribution behind shielding is negligible. The resulting primary spectra were then transported through \qty{10}{\gram\per\cm\squared} of aluminum, representative of a lightly shielded space habitat. All secondary particles (including neutrons, protons, and ions up to iron) with energies between \qty{1}{\MeV\per\atomicmassunit} and \qty{100}{\GeV\per\atomicmassunit} were scored. An additional underflow bin ensured that particles with energies below \qty{1}{\MeV\per\atomicmassunit} were also accounted for\replyTo{14}.  
Target spectra simulations were carried out using the General Particle Source (GPS) to define the spectral, spatial, and angular distributions of the source, together with the \texttt{QBBC-EMY} physics list, as recommended in the Geant4 official documentation for space radiation applications\replyTo{13}. The radiation field was generated within a spherical “source” volume enclosing a spherical mass model representing the shielding and “detector” geometries. An isotropic flux, representative of deep space, was modeled by sampling particle directions from a Lambertian distribution. To improve computational efficiency, source biasing was applied by restricting emission to a cone with a maximum half-aperture angle $\theta_{\text{max}}$ encompassing the entire geometry, following an approach similar to that described in \cite{peracchi2019modelling}. The shielding was implemented as a spherical \qty{10}{\gram\per\cm\squared} layer of aluminum alloy (6061, density \qty{2.7}{\gram\per\cm\cubed}).
The exposure time (t, in seconds), used for normalization, was determined from the number of simulated primaries ($N_{\text{s}}$) and the expected number of particles traversing the source volume in the real world ($N_{\text{r}}$):

\begin{equation}
    t = \frac{N_{\text{s}}}{N_{\text{r}}} = \frac{N_{\text{s}}}{\Phi \cdot 4 \pi^2 R^2 \sin^2(\theta_{\text{max}})} \quad [\si{\second}],
\label{eq:GCR_Target_spectra:t}
\end{equation}

where $\Phi$ is the all-ions, energy-integrated flux (assumed isotropic) from the GCR model, and the geometric factor accounts for the particle source volume. 
For a spherical detector of effective cross-sectional area $S_{\text{det}}$, the omni-directional differential flux (in \si{\count \per \centi\meter\squared \per \second \per \MeV \atomicmassunit}) was calculated as:

\begin{equation}
    \frac{d\phi}{dt dS_{\text{det}} dE} = \frac{C_{\text{s}}}{t \cdot S_{\text{det}}},
\label{eq:GCR_Target_spectra:dphi_over_dt_dS_dE}
\end{equation}

where $C_{\text{s}}$ is the number of particles scored per kinetic-energy bin (\unit{\MeV\per\atomicmassunit}). 
Using this procedure, differential kinetic energy spectra of GCR behind shielding were obtained. For later use in the optimization procedure, the spectra were restricted to the energy range from \qty{10}{\MeV\per\atomicmassunit} to \qty{2}{\GeV\per\atomicmassunit}.
The upper energy limit reflects the maximum primary-beam energy available at the GSI SIS-18 (\qty{1}{\GeV\per\atomicmassunit}) and the kinematic constraints of fragmentation processes, which confine heavier projectile-like fragments close to the beam energy while allowing only light fragments to extend up to about \qty{2}{\GeV\per\atomicmassunit}.
Corresponding Linear Energy Transfer (LET) distributions in water were derived by using the Bethe-Bloch equation with fitted free parameters on a precomputed LET($E$) table for each ion species.

\subsection{The GCR simulator at GSI: Development Strategy}
\label{ssec:GCRSimulator_at_GSI}

The GCR simulator at GSI is based on an active-passive approach combining energy switching of a single ion species ($^{56}$Fe) with passive beam modulation. Passive devices are of two types: (i) \textit{complex modulators}, similar to those used in hadron therapy, which mainly modulate the kinetic-energy distribution of heavy ions; and (ii) \textit{slab modulators}, which stop and fragment the primary beam to generate the light- and medium-$Z$ components of the GCR spectrum. 
Geometry, material composition, and thickness of all modulators are optimized so that a weighted combination of irradiations reproduces, within accelerator constraints, a predefined target GCR field.
The neutron component, inevitably produced by fragmentation in the passive modulators, additionally motivated the choice of simulating a shielded radiation scenario as discussed in \autoref{ssec:Galactic_Cosmic_Radiation_Environment}. In fact, neutrons are absent in the external free-space GCR field, and if shielding is artificially added downstream of the beamline, the modulators would still generate an additional neutron field dose not present in the original spectrum. By adopting the in-habitat reference case, this inconsistency is avoided, while the neutron component produced in the simulator is physically consistent with a shielded GCR environment. 
All simulations were carried out on GSI’s high-performance computing infrastructure using the Geant4 Monte Carlo toolkit version 11.2.1 \cite{Agostinelli2003, Allison2006, Allison2016}, with workload management via SLURM \cite{Yoo2003} and containerization through Apptainer \cite{gregory_m_kurtzer_2021_4667718}. The workflow follows a constrained optimization scheme, ensuring that results are both physically interpretable and experimentally feasible. 

\subsection{Basedata simulations}
\label{ssec:Basedata_simulations}
A “Basedata” set of simulations was generated by characterizing the radiation field emerging from different slabs of material, uniquely defined by: (i) their material composition, (ii) their thickness, and (iii) the primary beam energy. 
All simulations were performed using a $^{56}$Fe primary beam of user-defined energy and generated as a uniform spatial distribution across 
\begin{math}
x \in \left[-\qty{7.5}{\milli\metre}, \; \qty{7.5}{\milli\metre}\right]
\end{math}
and 
\begin{math}
y \in \left[-\qty{7.5}{\milli\metre}, \; \qty{7.5}{\milli\metre}\right]
\end{math} transverse plane, 
and directed along the longitudinal z-axis through a steel \textit{mesh modulator} designed to pre-scatter the beam and broaden its energy spread. 
The mesh modulator was constructed from stainless steel (Type: 304L) wires of \qty{60}{\micro \metre} diameter, arranged with a pitch of \qty{75}{\micro \metre} to form a half-layer. Two perpendicular half-layers were combined to form a full layer, and this process was repeated 32 times, with the layers placed in random orientation. After passing through the mesh, the beam interacted with a slab target (box-shaped, \qty{40}{\milli \metre} wide) of user-defined material and thickness, located at a fixed distance downstream of the mesh modulator.
Particles emerging from the target were scored at four distances (1000, 2000, 2500, and 3200 \si{mm} from the proximal target edge) for all fragments, ranging from neutrons ($Z = 0$) to iron ($Z = 26$). Scored quantities included (i) kinetic energy (in \qty{1}{\MeV\per\atomicmassunit} bins) and (ii) transverse position (x, y) at the scoring planes, modeled as thin (\qty{1}{\milli \metre}) cylinders with a radius of \qty{75}{\milli \metre} of air. 
To achieve well-formed kinetic energy distributions at \qty{1}{\MeV\per\atomicmassunit} resolution, a minimum of $\approx~10^6$ primary particles was required. For thicker materials, even larger numbers of primary particles were necessary to ensure statistical convergence. Simulations were performed for a discrete set of thicknesses. Intermediate thicknesses were estimated by linear interpolation, except in limiting cases such as very thin targets at high beam energy, or cases where the primary beam was almost fully stopped within the target. In such situations, additional simulations with finer thickness steps were performed. Interpolation was also avoided when new spectral features appeared (e.g., low-energy fragment peaks).
The main differences between Basedata for slab and complex modulators lie in the materials considered and the number of thicknesses required. In fact, manufacturing constraints strongly limit the range of materials available for complex modulators compared to slab targets. However, complex modulators require a much denser set of simulated thicknesses, since they must reproduce complex target functions with high spectral resolution. 
For slab modulators, an additional configuration combining a slab with a so-called “pin modulator” was investigated. This device, also referred to as FRAgments kiNetiC energy Optimizer (FRANCO), is designed to broaden the energy distribution of fragments emerging from slab targets. In this study, the pin modulator was implemented in polyethylene (PE), chosen for its efficiency in producing light fragments, and coupled with heavy slab materials capable of fully stopping the primary beam. Owing to the interplay of nuclear and electromagnetic interactions, slab–pin combinations allow both the required particle production and the required energy distributions to be achieved. 
Although FRANCO was not optimized with the same methodology as the complex modulators, it follows a similar geometrical design principle and is afterwards produced in a similar way. Each pin was modeled as a \qty{48}{\milli\metre}-high pyramid with a $2 \times 2$ \qty{}{\milli\metre\squared} square base, attached to a \qty{2}{\milli\metre} support plate, giving a total thickness of \qty{50}{\milli\metre} along a trajectory crossing the pin apex and base. The full modulator consisted of $20 \times 20$ repetitions of the unit pin structure, covering an area of $40 \times 40$ \qty{}{\milli\metre\squared}, and was integrated into the Monte Carlo geometry downstream of a slab target. 

\subsection{Design and Optimization of complex modulators}
\label{ssec:Design_and_Optimization_of_complex_modulators}

\begin{figure}[tbp]
    \centering
    \includegraphics[width=\columnwidth, keepaspectratio]{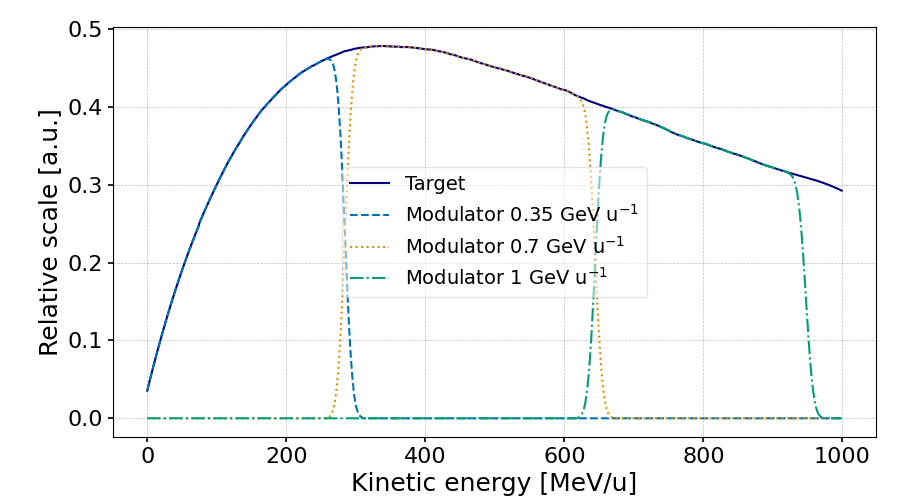}
    \caption{The target ($Z=26$) kinetic energy spectrum cannot be reproduced by a single modulator and was therefore subdivided to match the predefined primary beam energies of the accelerator. The high-energy region above \qty{1000}{\MeV\per\atomicmassunit} was excluded, corresponding to the accelerator's operational limit. Transitions between adjacent sections were smoothed using an error-function (ERF) to avoid sharp boundaries and minimize artifacts during the optimization.}
    \label{fig:Optimization_of_complex_modulators:Optimization_preparation}
\end{figure}

\begin{figure}[tbp]
    \centering
    \includegraphics[width=\columnwidth, height=10cm, keepaspectratio]{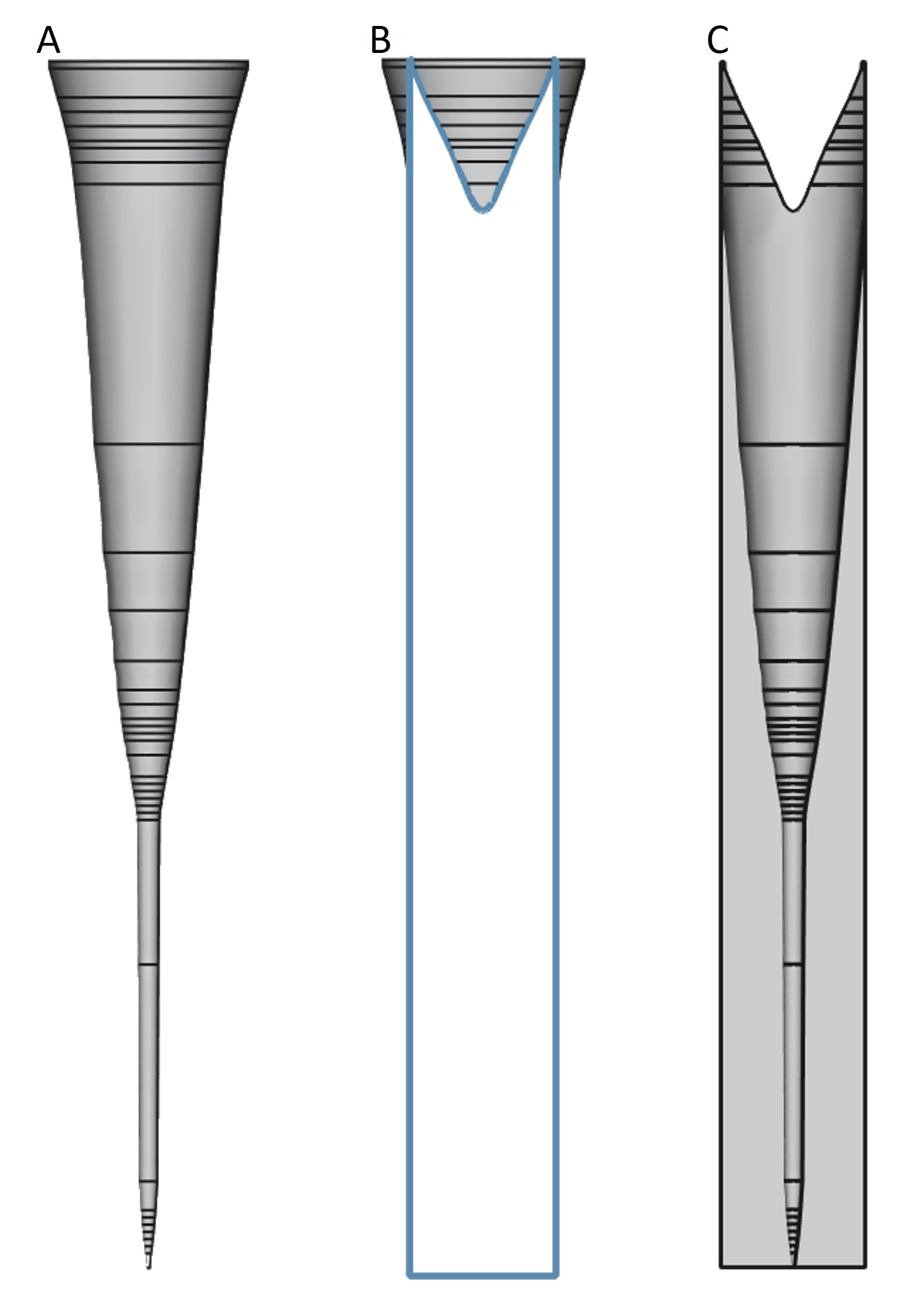}
    \caption{To generate the modulator structure for a single pin, the material weights are inverted and re-calculated to circular areas with different target thicknesses and lofted in FreeCAD to create a volume (Panel A). Afterwards, this volume is subtracted from a rectangular base structure (Panel B) with lateral dimensions of \qtyproduct{5 x 5}{\milli\meter} and an energy-dependent height (\autoref{fig:methods:MC_scheme_full_setup}), resulting in a complex hole and pin-like shape. The cross section of such a modulator structure is shown in Panel C. Depending on the desired lateral dimensions of the modulator, this basic modulation structure is arranged in a suitable sized matrix to reach the final dimensions.}
    \label{fig:Optimization_of_complex_modulators:Pin_generation}
\end{figure}

The design and optimization of complex modulators relies on a tool chain of several algorithms, heavily based on open-source software. ROOT 6.32.06 \cite{Brun1997} compiled with Minuit2 \cite{Hatlo2005} is used for data handling and weight optimization, while FreeCAD \cite{Riegel2025} converts weights into geometries. Several additional python modules, such as Uproot \cite{Pivarski2025}, NumPy \cite{Harris2020}, SciPy \cite{Virtanen2020} and matplotlib \cite{Hunter2007} interface the C++-based ROOT tools with the Python-based ones for tasks such as post-processing target spectra. 
In a first step, the target simulations (\autoref{ssec:Galactic_Cosmic_Radiation_Environment}) were post-processed to facilitate the optimization process. To avoid oscillation of the optimizer, the spectra were smoothed. The kinetic energies of the target spectra had to match the three corresponding primary beam energies. Hard cuts were not suitable for this separation, so an error (ERF) function was applied (\autoref{fig:Optimization_of_complex_modulators:Optimization_preparation}). 
The optimization iteratively varies the relative material weights of the provided sets of Basedata (\autoref{ssec:Basedata_simulations}) per primary beam energy to reach the target spectrum using the three complex modulators. 
In practice, since the number of weights is high, a subset of nodes is used for direct optimization. Nodes represent weights employed directly in the minimization process, while non-node weights are linearly interpolated based on surrounding nodes. This approach reduces computational load and mitigates oscillations in the calculated weights. Additional constraints are applied to the calculated weights, including general limits and parameters of the minimization engine, such as the minimum step size, to ensure that the modulator can be converted into a producible geometry. 
In the workflow, the number of nodes was increased between consecutive runs of the optimizer on the same datasets to refine the results. The optimized weight distributions were then converted into a suitable production geometry using a custom FreeCAD macro that generates individual complex pin and/or hole modulation structures based on the weights and user-provided information. A detailed description of the geometry conversion process is given in \cite{Pfuhl2024}; here it is briefly summarized.
As shown in \autoref{fig:Optimization_of_complex_modulators:Pin_generation} for the \qty{0.35}{\GeV\per\atomicmassunit} complex modulator, the modulation structure is generated by converting the weights to areas, which are then lofted and subtracted from the modulator base. The base structure is quadratic, while the optimized volume, which is removed from the base, is circular. Circular areas are recalculated, as described in \cite{Pfuhl2024}, providing higher mechanical rigidity, ease of production, and complete area coverage without non-optimized material. 
Each modulation structure (pin/hole) is converted to a mesh, saved as stereolithography (STL) file, and imported into a Geant4 simulation via CADMesh \cite{Poole2012, Poole2012a}. Using only the STL of a single pin/hole structure and multiplying it allows for more computationally efficient simulations. Simulations with the resulting optimized complex modulators are then performed at the appropriate energies to verify that the target spectra are accurately reproduced. 

\subsection{GCR simulator optimization process}
\label{ssec:GCRSimulator_Optimization_process}
\replyTo{5}To ensure an implementable and experimentally feasible design of the GCR simulator components, the optimization process accounted for several constraints, including the maximum deliverable energy from the accelerator, the limited space on the beamline, the maximum weight that can be handled by the automated modulator exchangers. In addition, to keep the irradiation time within acceptable limits, a total of six configurations was selected: three defined by complex modulators and three based on slab modulators, whose relative contributions were treated as free parameters to be determined by the optimization. The objective is to reproduce the target kinetic-energy spectra for all relevant ion species ($Z \in [1, 26]$). Neutrons ($Z = 0$), although part of the space radiation field, were excluded from this step, as directly optimizing their spectral distribution would conflict with the physical processes, and consequently the configurations, required to produce and tune the charged-particle components of the GCR simulator. Furthermore, neutron production and transport are subject to larger model-dependent uncertainties and rely on more limited experimental data than fragmentation processes involving charged particles.
The optimization determines a set of six weights $\omega$, one for each configuration, that minimize the discrepancy between the weighted combination of simulated spectra and the corresponding target spectra. The procedure is performed directly on kinetic-energy distributions, which fully characterize the radiation field in terms of particle type, abundance, and energy. An optimization carried out solely in LET or dose would be overly simplistic, since particles of different charge and energy can produce the same LET value; such an approach could reproduce the LET distribution while failing to match the underlying particle composition and energy spectra characteristic of the GCR environment.
All mathematical formulations, including the cost function, the spectral matrices, and the complete optimization formalism, together with the software implementation, are provided in \ref{app:simulator_optimization}.

\subsection{Dosimetric quantities calculation}
\label{ssec:Dosimetric_quantities_calculation}
\replyTo{17}In space radiation protection, the assessment of stochastic risks, such as carcinogenesis, cannot be described by absorbed dose alone, due to mixed field of high-energy protons, neutrons, and heavy ions that differ markedly in their relative biological effectiveness (RBE) \cite{Furukawa2020}. To account for these differences, the dose equivalent is introduced as the product of absorbed dose and a radiation quality factor, $Q$. Two main approaches are currently used. The ICRP definition \cite{ICRP60} expresses $Q$ solely as a function of LET in water, assuming that biological effectiveness scales with the local density of energy deposition along particle tracks. Conversely, NASA \cite{cucinotta2013space} introduced a formulation where $Q$ depends not only on LET but also explicitly on particle charge and velocity through the parameter $Z^{*\,2}/\beta^2$. This allows a more accurate description of track-structure effects, particularly for heavy ions. Accordingly, for ions with charge \qtyrange{1}{26}{} and kinetic energy from \qtyrange{10}{2000}{\MeV\per\atomicmassunit}, the dose equivalent was computed either from the LET fluence spectrum (ICRP formalism), or from kinetic energy fluence spectra (NASA approach). The dose-averaged quality factors were then calculated as:

\begin{align}
& \overline{Q}_{\mathrm{ICRP}} = \frac{\int Q(\mathrm{LET}) \, D(\mathrm{LET}) \, d\mathrm{LET}}{\int D(\mathrm{LET}) \, d\mathrm{LET}},
\label{eq:ICRP_Q}
\end{align}

\begin{align}
& \overline{Q}_{\mathrm{NASA}} = \frac{\sum_Z \int Q(Z,E)\, D(Z,E)\, dE}{\sum_Z \int D(Z,E)\, dE}.
\label{eq:NASA_Q}
\end{align}

While quality factor functions are well established for cancer risk estimation, only preliminary RBE-based risk estimates exist for other endpoints such as cardiovascular disease (CVD) \cite{cucinotta2013safe}, and experimental data from rodent studies provide limited insight into potential central nervous system (CNS) effects \cite{kiffer2019behavioral, liu2019space, cucinotta2019risks}. In this work, quality factor calculations follow the currently accepted formulations for cancer risk, providing a consistent quantitative framework widely used in space radiation research. Since the development and optimization of the GCR simulator were based solely on the underlying physical properties of the GCR environment, without incorporating biological weighting or specific endpoint assumptions, the simulator is readily applicable to a broad range of radiobiological investigations beyond cancer.

\section{Results}
\label{sec:Results}

\subsection{Monte Carlo simulations}
\label{ssec:Monte_Carlo_simulations}
\begin{figure*}[tbp]
    \centering
    \includegraphics[width=\textwidth]{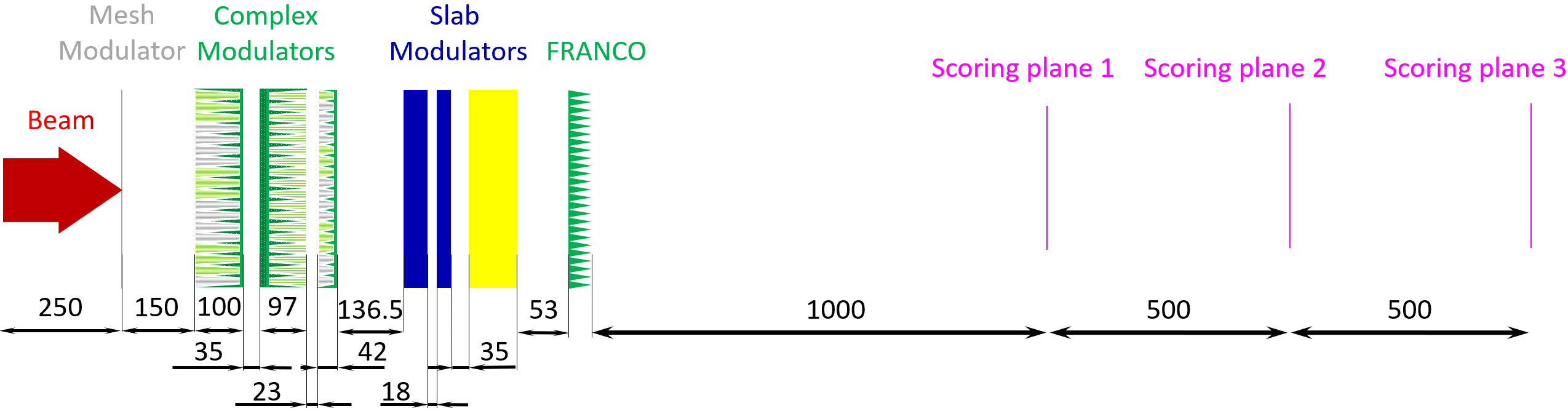}
    \caption{Schematic representation of the GCR simulator along the beamline, implemented in Geant4, including all components and their relative distances. The beam is generated from a squared spatial distribution with $x \in \left[-40,\, 40 \right] \, \unit{\mm}$ and $y \in \left[-40,\, 40 \right] \, \unit{\mm}$, orthogonal to the beam direction. The first element along the beam path is the steel mesh modulator, followed by complex modulators designed for primary beam energies of \qty{1}{\GeV\per\atomicmassunit}, \qty{0.7}{\GeV\per\atomicmassunit}, and \qty{0.35}{\GeV\per\atomicmassunit} (shown in green), slab modulators (two steel-304L slabs in blue and a polyethylene (PE) slab in yellow), and the FRANCO modulator. Except for the steel mesh modulator, which is permanently installed, all modulators are positioned along the beamline according to the specific configuration defined for each exposure. Three scoring planes are placed downstream of the FRANCO modulator to evaluate the spatial homogeneity of the radiation field. All dimensions in the sketch are given in \unit{\mm}.}
    \label{fig:methods:MC_scheme_full_setup}
\end{figure*}

The optimization software described in \autoref{ssec:Design_and_Optimization_of_complex_modulators} and \autoref{ssec:GCRSimulator_Optimization_process} produced a set of configurations, coupling primary beam energies and modulators, capable of reproducing the target radiation field, namely the GCR behind \qty{10}{\gram\per\cm\squared} Al at \qty{1}{\astronomicalunit} for 2010 solar minimum. As noted in \autoref{ssec:Basedata_simulations}, all configurations reported below incorporate a 32-layer steel mesh modulator. Once the six suitable configurations were identified, new Monte Carlo simulations were performed to develop feasible setups with finalized geometries which are presented in \autoref{tab:config_weights}. Additionally, \autoref{fig:methods:MC_scheme_full_setup} illustrates the selected implementation of GSI's hybrid active-passive GCR simulator, including the configurations and the inter-element distances necessary to reproduce the target GCR field.\\ 
The first beamline element, the 32 layer stainless steel mesh modulator described in \autoref{ssec:Basedata_simulations}, is followed by the three optimized complex modulators. These were implemented as mesh geometries using CADMesh, in Visijet M2S-HT250, with a density of \qty{1.1819}{\gram \per \cm \cubed}. Slab modulators were modeled as box-shaped elements (\qtyproduct{100 x 100}{\mm}) made either of Steel-304L or polyethylene (PE, via the NIST material database).
Immediately after the slab modulators, FRANCO with a lateral size of \qtyproduct{100 x 100}{\mm} was added, also implemented in M2S-HT250. The system was optimized to reproduce the GCR
\qty{2000}{\milli \metre} after the distal edge of the last beamline element (FRANCO) in a \qtyproduct{75 x 75}{\mm} area.

\begin{table}[htbp]
\centering
\footnotesize
\caption{Optimized GCR simulator configurations and corresponding weights.}
\label{tab:config_weights}
\begin{tabularx}{\columnwidth}{L >{\raggedright\arraybackslash}S[table-format=3.3e0]}
\toprule
\textbf{Configuration} & {\textbf{Weights} (\unit{\per\second})} \\
\midrule
Complex modulator \qty{1}{\GeV\per\atomicmassunit} & \num{9.656e-3} \\
\midrule
Complex modulator \qty{0.7}{\GeV\per\atomicmassunit} & \num{1.322e-2} \\
\midrule
Complex modulator \qty{0.35}{\GeV\per\atomicmassunit} & \num{4.718e-3} \\
\midrule
Slab modulator \qty{0.35}{\GeV\per\atomicmassunit},\\
\qty{50}{\milli\metre} Steel + \qty{100}{\milli\metre} PE & \num{3.621e4} \\
\midrule
Slab modulator \qty{1}{\GeV\per\atomicmassunit},\\ 
\qty{80}{\milli\metre} Steel + FRANCO & \num{3.944e2} \\
\midrule
Slab modulator \qty{1}{\GeV\per\atomicmassunit},\\ 
\qty{50}{\milli\metre} Steel + FRANCO & \num{1.002} \\
\bottomrule
\end{tabularx}
\end{table}

\begin{figure*}[tbp]
    \centering
    \includegraphics[width=0.8\textwidth]{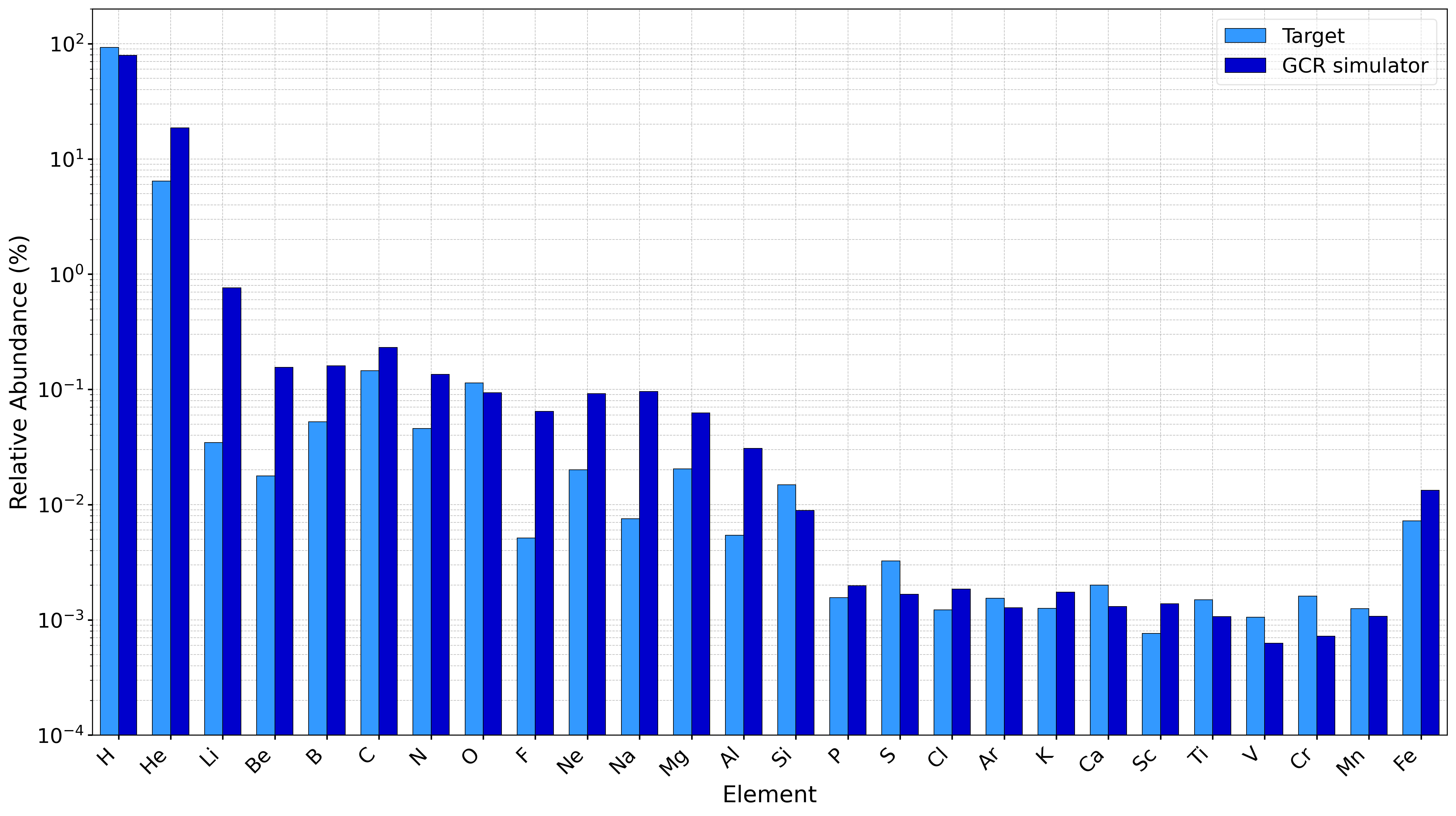}
    \caption{Relative elemental abundances in the reference GCR spectrum (light blue) compared with those generated by the GCR simulator (dark blue). The comparison highlights the capability of simulator to reproduce the elemental composition of the galactic cosmic ray environment across a wide range of nuclear species.}
    \label{fig:Results:Abundance}
\end{figure*}

\begin{figure}[tbp]
    \centering
    \includegraphics[width=\columnwidth]{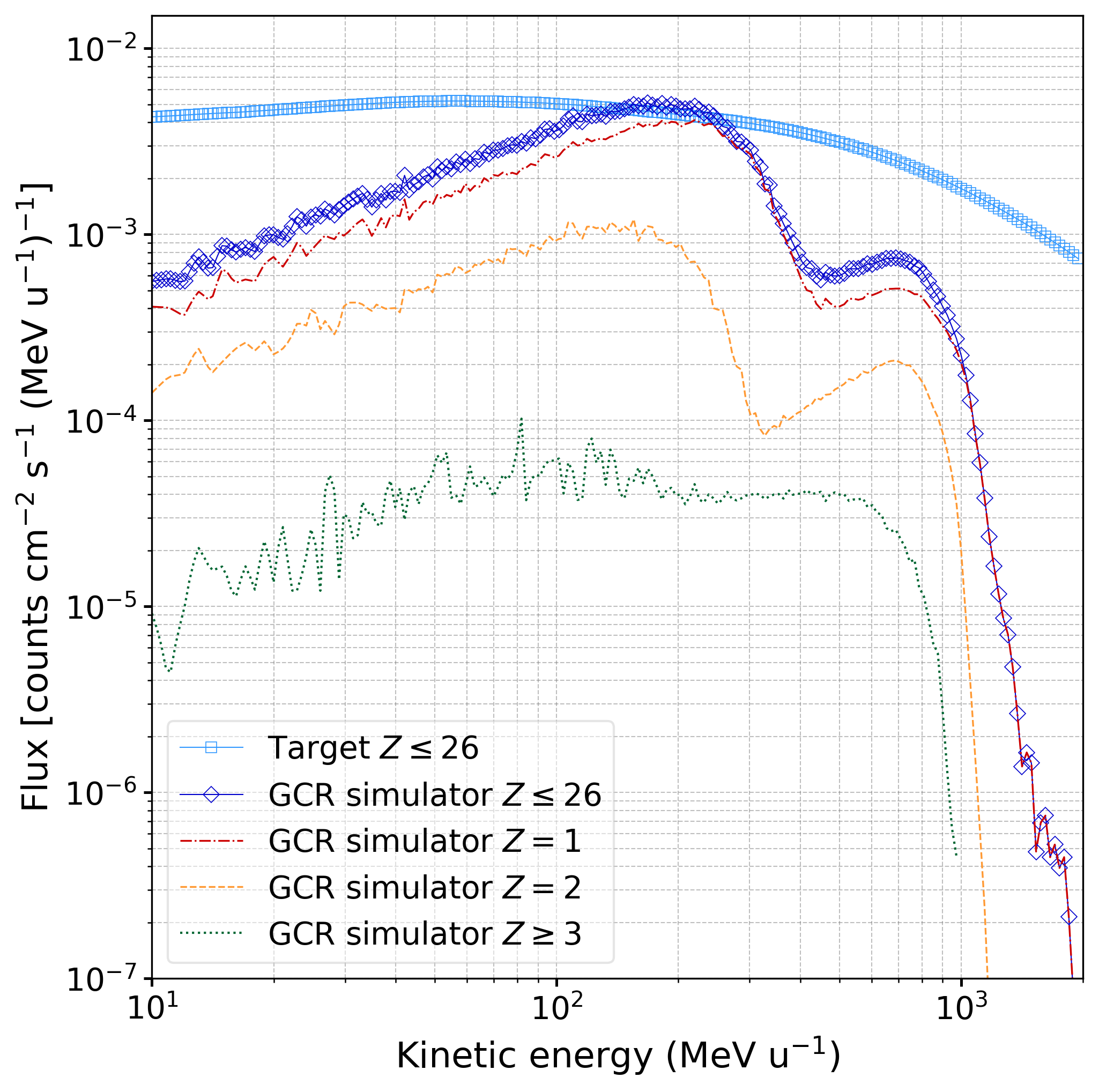}
    \caption{Differential kinetic energy spectra of the reference GCR (light blue) and of the GCR simulator (dark blue), both obtained by summing over all ion species ($Z\leq26$). For the GCR simulator, the separate contributions of protons ($Z=1$, red), helium nuclei ($Z=2$, orange), and HZE nuclei ($Z \geq 3$, dark green) are also displayed.}
    \label{fig:Results:Kinetic_Energy}
\end{figure}

\begin{figure*}[tbp]
    \centering
    \includegraphics[width=0.8\textwidth]{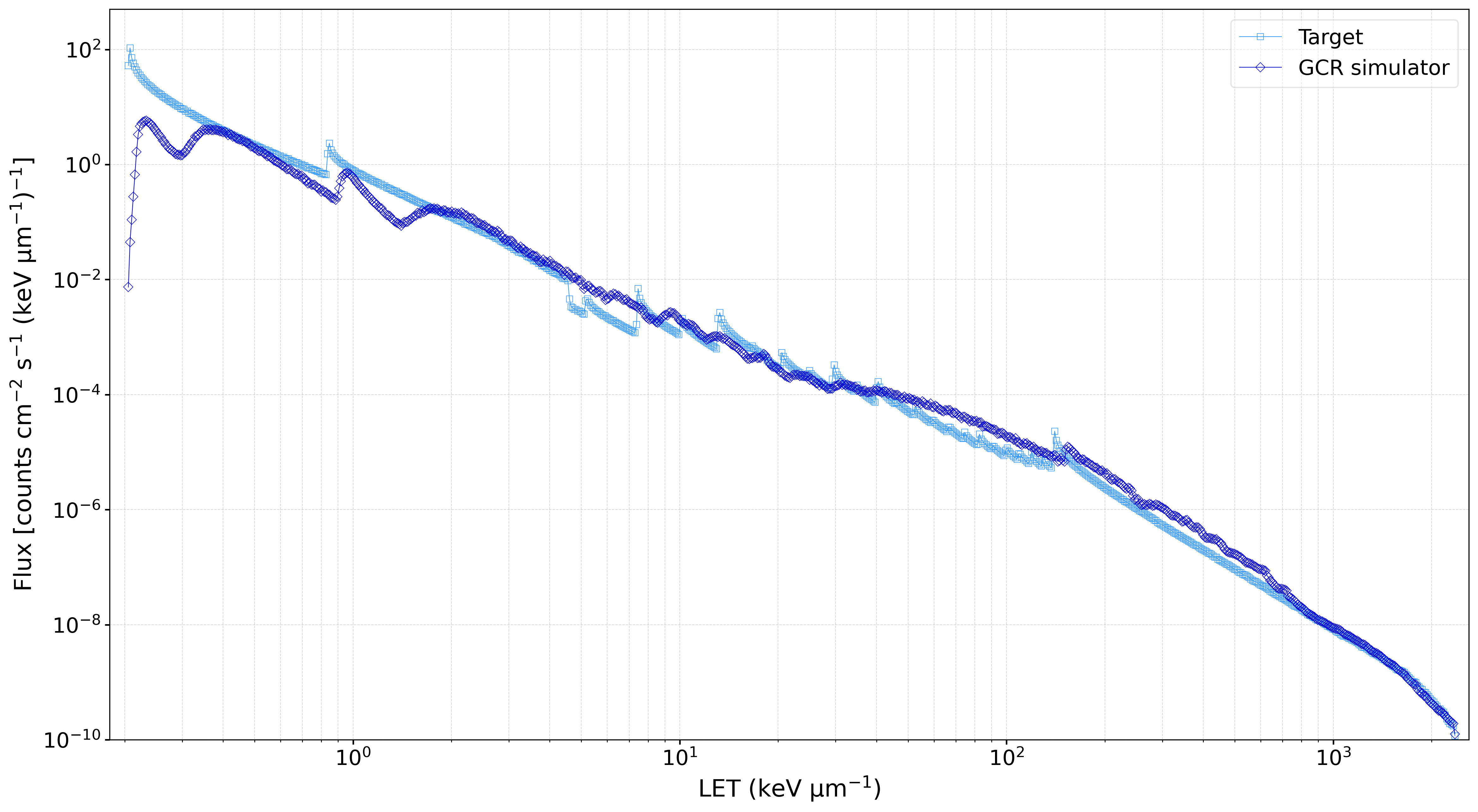}
    \caption{Linear energy transfer (LET) spectra of the reference GCR (light blue) and of the GCR simulator output (dark blue).}
    \label{fig:Results:LET}
\end{figure*}

The first three weights, corresponding to complex modulators, are of the same order of magnitude, whereas the weights assigned to the slab modulators are significantly higher and less uniform. $\omega$ represents the number of primary particles per unit of time interacting with a given configuration and, therefore, higher particle numbers need to be assigned to configurations responsible for producing lighter ions with higher abundance. Additionally, slab modulators exhibit a lower conversion frequency compared to complex modulators as they need to fully stop and fragment the primary beam.
By applying these weights to the combined setup described above, a GCR-like radiation field can be generated and all relevant spectra can be calculated. These are presented in \autoref{fig:Results:Abundance}, \autoref{fig:Results:Kinetic_Energy} and \autoref{fig:Results:LET}, and are compared to the corresponding optimization targets. \\
\replyTo{28}\autoref{fig:Results:Abundance} shows the relative abundances of the relevant elements (from H to Fe) produced by the GCR simulator in the energy range from \qtyrange{10}{2000}{\MeV\per\atomicmassunit}, compared with their respective target distributions.
Quantitatively, approximately $50\%$ of the elements are reproduced within a factor of 2, and $73\%$ within a factor of 3, with a median log-ratio corresponding to a typical deviation of roughly $30\%$ above the expected values. The over-representation of iron and lighter-to-intermediate ions (He–Al) relative to the target reflects the inherent limitations in reproducing a mixed field using a single-ion primary beam combined with secondary fragmentation products. Despite these deviations, the simulator reproduces the full range of ionic species (H–Fe), including neutrons (not shown in \autoref{fig:Results:Abundance}), and captures the cumulative contribution of all species, ensuring a representative ground-based analog of the GCR environment.
\replyTo{20}The total kinetic-energy spectrum, summing particles from protons to iron within the considered energy interval, is presented in \autoref{fig:Results:Kinetic_Energy}. The trend observed in the abundance plot is confirmed, with protons and helium ions dominating the total differential flux.
Agreement with the target spectrum is generally satisfactory in the intermediate energy region (several tens of \qty{}{\MeV\per\atomicmassunit} to about \qty{1000}{\MeV\per\atomicmassunit}). In this range, the GCR simulator spectrum presents a characteristic double-peak structure, in contrast with the single, broadened peak of the target spectrum. This arises from the slab-modulator configurations used to stop the primary ${^{56}}$Fe beam and produce low-$Z$ fragments. Since the fragment kinetic energies cluster around the nominal primary beam energies (\qty{0.35}{\GeV\per\atomicmassunit} and \qty{1}{\GeV\per\atomicmassunit}), the resulting spectrum features two distinct maxima rather than a smooth distribution. The pin-modulator device (FRANCO) broadens the fragment kinetic-energy distributions but cannot fully compensate for the discrete set of available primary beam energies.
Above approximately \qty{1000}{\MeV\per\atomicmassunit}, the progressive loss of accuracy relative to the target spectrum reflects the intrinsic limitation imposed by the maximum nominal energy of the primary beam, as discussed earlier. Achieving a smoother and more continuous kinetic-energy distribution would require additional slab-modulator configurations with more finely spaced primary beam energies, at the cost of increased simulator complexity and longer irradiation times.
For completeness, kinetic-energy spectra of all particle species, from neutrons ($Z=0$) to iron ($Z=26$), comparing Target and GCR simulator, are shown in \ref{app:kinetic_energy_spectra}.
The LET spectrum (\autoref{fig:Results:LET}) exhibits a generally close match across the entire range of interest (\qtyrange{0.2}{1000}{\keV\per\micro\metre}) between the target distribution and the optimized GCR simulator field. It is important to note that LET was not a direct optimization target but serves as a proxy to assess the mixed-field quality resulting from the optimized kinetic-energy spectra. Because LET is fully determined by the underlying particle species and their kinetic-energy distributions, the observed agreement in \autoref{fig:Results:LET} reflects the consistency of the optimized energy spectra rather than any direct tuning performed in LET.
Nevertheless, the LET spectrum is presented because it is a widely adopted experimental observable and a standard descriptor of radiation-field quality, although alternative parameters such as $Z^{*\,2} / \beta^2$ may provide a more detailed characterization of microscopic energy depositions, particularly for heavy ions.
Consistent with the behaviour observed in the kinetic-energy spectra, the LET distribution exhibits characteristic structures associated with individual ions. In the low-LET region, the two peaks at approximately \qty{0.2}{\keV\per\micro\metre} and \qty{0.85}{\keV\per\micro\metre}, correspond to protons and helium ions, respectively. Their relative heights and widths reflect the efficiency with which low-$Z$ particles are produced by the simulator compared to the target field. At higher LET, a peak near \qty{160}{\keV\per\micro\metre} is present, arising from the dominant contribution of iron ions, which shape the high-LET region in both distributions.
\replyTo{21}As outlined in \autoref{ssec:Dosimetric_quantities_calculation}, cancer risk from exposure to complex radiation fields is commonly assessed using quality factors. \autoref{tab:config_quality_factors} reports the dose-averaged quality factors calculated for each individual GCR simulator configurations, as well as for the full simulator and the reference target field, according to \autoref{eq:ICRP_Q} and \autoref{eq:NASA_Q} for the ICRP and NASA definitions, respectively.
\replyTo{22}In line with the LET distribution shown in \autoref{fig:Results:LET}, the GCR simulator exhibits higher dose-averaged quality factors than the target field under both definitions. This outcome reflects the particle yield of the simulator: the slab-modulator configurations enhance the production of intermediate-$Z$ fragments within specific kinetic-energy intervals, increasing the high-LET component of the field. Simultaneously, the relative abundance of low-$Z$ particles, particularly protons, which would otherwise reduce the dose-averaged quality factor, is decreased. These combined effects result in an overall increase of $\overline{Q}$ values for the full simulator. It is important to emphasize that the GCR simulator aims to realistically approximate the space radiation environment within operational and experimental constraints, although it is not expected to replicate every physical descriptor exactly. Furthermore, the dose-averaged quality factor is a biological effectiveness metric specifically associated with cancer risk. Other health endpoints of interest (e.g., cardiovascular or central nervous system effects) would rely on distinct weighting functions, and the degree of agreement between simulator and target may vary depending on the endpoint considered.

\begin{table}[htbp]
\centering
\footnotesize
\caption{Optimized configurations and corresponding radiation quality factors. Errors represent the standard deviation obtained from Monte Carlo bootstrapping/resampling of the simulation results following a similar procedure also used in \cite{george2024space}.}
\label{tab:config_quality_factors}
\begin{tabularx}{\columnwidth}{L  S[table-format=1.3] S[table-format=1.3]}
\toprule
\textbf{Configuration} & $\overline{Q}_{\mathrm{ICRP}}$ & $\overline{Q}_{\mathrm{NASA}}$\\
\midrule
Complex modulator \qty{1}{\GeV\per\atomicmassunit} & \num{24.015 \pm 0.001} & \num{24.550 \pm 0.001} \\
\midrule
Complex modulator \qty{0.7}{\GeV\per\atomicmassunit} & \num{21.743 \pm 0.001} & \num{25.754 \pm 0.001 } \\
\midrule
Complex modulator \qty{0.35}{\GeV\per\atomicmassunit} & \num{13.786 \pm 0.002} & \num{14.864 \pm 0.004} \\
\midrule
Slab modulator \qty{0.35}{\GeV\per\atomicmassunit},\\ \qty{50}{\milli\metre} Steel + \qty{100}{\milli\metre} PE & \num{1.050 \pm 0.003} & \num{1.076 \pm 0.003} \\
\midrule
Slab modulator \qty{1}{\GeV\per\atomicmassunit},\\ \qty{80}{\milli\metre} Steel + FRANCO & \num{10.536 \pm 0.028} & \num{12.177 \pm 0.034} \\
\midrule
Slab modulator \qty{1}{\GeV\per\atomicmassunit},\\ \qty{50}{\milli\metre} Steel + FRANCO & \num{18.153 \pm 0.009} & \num{19.946 \pm 0.013} \\
\bottomrule
\toprule
\multicolumn{1}{l}{\textbf{GCR simulator}} & \num{5.60 \pm 0.02} & \num{6.33 \pm 0.02} \\
\midrule
\multicolumn{1}{l}{\textbf{Target (2010 Sol Min)}} & \num{3.24} & \num{3.36} \\
\bottomrule
\end{tabularx}
\end{table}

\section{Discussion}
\replyTo{2, 3, 4, 23}NASA’s active GCR simulator was developed to reproduce the radiation environment at the female blood-forming organs behind \qty{20}{\gram\per\cm\squared} aluminum during solar minimum, using thirty-three sequential mono-energetic beams of proton and helium (with degraders extending their spectrum below \qty{100}{\MeV\per\atomicmassunit}), and five representative heavy ions. However, the step-wise approach exposes samples to a sequence of narrow energy and LET distributions instead of the complex mixed radiation field encountered in space.
The hybrid active–passive concept adopts a fundamentally different strategy. Each of the six configurations produces a broad particle spectrum, spanning multiple charges and kinetic energies simultaneously through the fragmentation and slowing down of a primary $^{56}$Fe beam in passive modulators. Mission-dependent heliospheric conditions can then be reproduced via software by re-weighting these six configurations without hardware changes, providing a continuous and realistic LET distribution relevant for dose-equivalent estimates while preserving flexibility for physics, biology, and shielding studies.
A notable limitation of the NASA GCR simulator is the absence of a neutron component, despite neutrons being copiously produced when primary GCR ions interact with spacecraft materials and human tissue. These energetic secondaries are an important contributor to astronaut exposure, generating recoil protons and light charged fragments ($Z \leq 2$), and, via interactions with nuclei in the human body, target fragments with $Z>2$ and high LET. Such tissue-target fragments, associated with potential biological risk, are not represented in the NASA reference field \cite{huff2023galactic, norbury2016galactic}.
In contrast, the hybrid approach naturally generates a broad neutron spectrum (\ref{app:kinetic_energy_spectra}) through the fragmentation of both primary beam and the modulators. Although the relative neutron abundance exceeds that of the reference field, using kinetic-energy-dependent RBE functions for Double Strand Break (DSB)-cluster induction \cite{mentana2025mapping}, the RBE-weighted neutron contribution is estimated to be $\sim 39\% $ lower than for the reference spectrum. Thus, despite quantitative differences and associated simulation uncertainties, the resulting field includes a neutron component that is absent, or only partially represented, in sequential-beam approaches, thereby enabling dedicated studies of its biological impact.
Moreover, neutron interactions are stochastically distributed throughout the body in space, generating high-LET events due to the multiplicity of neutron-induced target fragments \cite{huff2023galactic}. The NASA sequential delivery does not preserve these spatial correlations, whereas the mixed, simultaneous field, delivered by the hybrid simulator retains them intrinsically. Likewise, the sequential nature of the NASA simulator limits its ability to reproduce the spatial-temporal correlations arising from the interaction of heavy GCR ions with shielding or tissue, where multiple projectile-like fragments may traverse the same microscopic volume nearly simultaneously. Preservation of such correlations may be relevant for CNS-related endpoints, where distributed and near-synchronous track structures could play a role.
While the GSI approach reproduces key GCR field features (neutrons, spatial-temporal correlations, and broad spectral distributions), it also exhibits its own limitations. The use of a single primary beam produces stray radiation and secondary components requiring careful characterization, and field uniformity is only guaranteed within a defined geometry (\qtyproduct{7.5 x 7.5}{\cm} at \qty{2000}{\mm} from the last element), constraining target positioning. Nonetheless, optimizing a GCR-like field directly behind light-to-moderate shielding (\qty{10}{\gram\per\cm\squared}) allows additional material to be placed upstream of the target and enables direct testing of novel shielding concepts under space-analog conditions.
Beyond these aspects, biological considerations introduce further complexity.
Radiobiological studies\replyTo{24} \cite{zhou2006proton, mcnally1984effect, ngo1981sequential, durand1976irradiation, hada2007chromosome, hada2021chromosome, huang2019synergy, huang2020simulating} suggest that the order and timing of irradiation influence outcomes, as different biological responses may be triggered depending on the sequence and the time interval between exposures.
Estimates indicate that in space\replyTo{25} every cell in an astronaut's body is traversed by a proton every few days, by helium nuclei every few weeks, and by HZE particles only every few months \cite{Durante2011}. The traversals are neither random nor statistically independent, as the passage through a nucleus typically coincides with the traversal of approximately \qty{1e9}{} additional nuclei along the same track. Thus, absorbed dose or dose rate alone
are misleading descriptors, as energy deposition is highly heterogeneous both physically and temporally, and biological effects depend strongly on the endpoint considered \cite{norbury2016galactic}.
From a probabilistic perspective, cells are more likely to be hit by protons and helium ions before HZE particles, suggesting that simulator configurations delivering light ions should ideally precede, or being interleaved with, those producing the heavy ion component. Practical constraints, however, must also be considered:
configurations sharing the same primary iron beam energy should be grouped together to minimize accelerator energy changes and limit accelerator overhead time. These aspects highlight the need for further theoretical and experimental work to clarify the biological impact of irradiation in mixed-field exposures, and to establish the most appropriate sequence to reproduce the space environment. Addressing this issue is crucial not only to predict biological outcomes, but also to ensure comparability across experiments performed with different GCR simulator concepts.
At GSI, the Radiobiology Modelling group is developing a dedicated simulation framework based on the Local Effect Model (LEM) to explore and quantify these effects \cite{Pfuhl2019}.\\ 
Operational considerations are also crucial. For context, the NASA beam sequence requires approximately one hour to deliver the full setup. In the hybrid system, the interval between configurations is currently limited by accelerator-energy switching and modulator exchange to $\lesssim$~\qty{1}{\minute}. Ongoing developments aim to reduce this interval to the inter-spill time (\qtyrange{1}{2}{\second}), at which point the delivery of the six weighted configurations would effectively constitute a quasi-continuous irradiation, closely mimicking chronic space exposure. As an estimate, for a Mars-mission reference dose of \qty{300}{\milli\gray} over 650 days, the GSI's GCR simulator can deliver the equivalent mixed-field exposure in $\lesssim$~\qty{30}{\minute}, and \qty{500}{\milli\gray} in $\lesssim$~\qty{1}{\hour}, covering particles from neutrons to iron over the kinetic-energy range \qtyrange{10}{2000}{\MeV\per\atomicmassunit}. This calculation assumes technical parameters of GSI's experimental vault (Cave A) and SIS-18, including a typical particle rate of \qty{1e9}{\primary\per\second} and a duty cycle of \qty{10}{\second}, with \qty{8}{\second} beam-on and \qty{2}{\second} spill pause. 

To facilitate in-silico user studies and reduce computational time, a user-defined particle source for Geant4 was generated to reproduce the complex radiation field of the hybrid active-passive GCR simulator without the need to implement the full setup geometry or transport the primary beam. This enables users to bypass the computationally intensive step of simulating the entire beamline and instead focus directly on their specific research questions. This functionality is implemented in the tool G4\_\-PS\_\-converter, which generates a Phase Space for Geant4 via sets of appropriate macro (.mac) commands. These macros contain all relevant quantities, kinetic energy, abundance, position ($x,y$), and azimuthal and polar angles ($\theta, \phi$), for each particle species, derived from Geant4 simulations of the complete GCR simulator setup performed using GSI's cluster computing resources. Geant4 can then use these macros to accurately regenerate the GCR simulator radiation field in any custom simulation according to the needs of the user. To streamline user interactions, a dedicated macro file is also provided, containing the instructions necessary to execute all the generated .mac commands in sequence.\\
The G4\_PS\_converter is publicly available at \url{https://github.com/chrischu0815/g4_ps_converter}.

\section{Conclusions}

This work presents the design and development of the hybrid active-passive GCR simulator implemented in GSI's experimental vault, Cave A. The system implementation and benchmarking through microdosimetric measurements are detailed in \cite{pierobon2025exp}. Our approach uses a single ion species ($^{56}$Fe), at three different primary beam energies, interacting with a total of six passive complex or slab modulators. The superimposition of six optimized irradiation configurations reproduces a GCR-like field representative of conditions at \qty{1}{\astronomicalunit} behind \qty{10}{\gram\per\cm\squared} of aluminium shielding during quiet solar activity (2010 solar minimum). 
The GCR field after \qty{10}{\gram\per\cm\squared} of aluminium shielding constitutes the baseline on which the optimization of the simulator was performed. Building on this baseline, the system remains intrinsically flexible: additional physical material can be placed along the beamline to investigate alternative shielding architectures or specific biological endpoints, and different solar modulation conditions can be reproduced by software-based re-tuning of the weights of the six configurations. This flexibility enables the GSI's GCR simulator to support a wide range of experimental needs in space radiation effects research.
The optimization process of the passive modulators, together with the scaling factors required to reproduce a GCR-like radiation field, is described in detail in this work. Comprehensive in-silico benchmarking of the GCR simulator, including the LET spectrum and quality factor estimations, is also provided. Finally, a dedicated Geant4-based particle source, the Phase Space, has been developed to support simulation studies and the planning of future experiments. This resource is openly available to the community and can be accessed at \url{https://github.com/chrischu0815/g4_ps_converter}.

\section*{Acknowledgements}
The development of the GCR simulator is supported by ESA under contract number 4000102355\allowbreak/10/NL/VJ. We would also like to acknowledge the additional support by the EU project HEARTS, a project funded by the European Union under GA 101082402, through the Space Work Programme of the European Commission.
This research was partly supported  by the cluster computing resource provided by the IT Department at the GSI Helmholtzzentrum f{\"{u}}r Schwerionenforschung, Darmstadt, Germany.
The current work was performed at the GSI Helmholtzzentrum f{\"{u}}r Schwerionenforschung in Darmstadt (Germany) within the frame of FAIR Phase-0.

\bibliographystyle{elsarticle-num}
\bibliography{references_paper_V2}

\newpage
\appendix
\section{GCR simulator optimization process}
\label{app:simulator_optimization}
To reproduce the reference target spectra under the experimental constraints described in \autoref{ssec:GCRSimulator_Optimization_process}, an optimization procedure was formulated to identify a set of weights, one for each of the six available simulation configurations, that minimizes the difference between their weighted sum (including the already optimized complex modulators) and the corresponding target distributions. The comparison is performed bin-wise in the kinetic energy spectra for each ion species.
Neutrons ($Z=0$) are excluded from the determination of these weights to limit the propagation of model-related uncertainties. In Monte Carlo transport, neutron production and neutron–nucleus interactions carry significantly larger systematic uncertainties than the charged component, due to limitations in neutron cross-section databases and model dependence in the relevant energy ranges. Including neutrons in the optimization would also introduce an intrinsic inconsistency with the nuclear and electromagnetic interactions required to reproduce charged-particle spectra, since matching the space-like neutron field would necessitate configurations whose characteristics compete with those needed to accurately simulate the charged component.
For this reason, the optimization relies exclusively on the charged particles, for which transport models are more strongly constrained by experimental data. Importantly, once the optimal weight vector is obtained, it is applied consistently to all particle species, including neutrons, when evaluating the final mixed field, ensuring that the neutron contribution is still represented in the overall mixed field without biasing the optimization process.

Formally, let the index $i$ denote the $i$-th configuration ($i \in [0, 5]$) and the index $Z$ denote the charge number ($Z \in [0, 26]$). For each $(i, Z)$ pair, the differential kinetic energy spectrum can be expressed as:
\[
\widetilde{E}_{i,Z} = \left\{ \mathfrak{B}_{i, Z}, \, \mathfrak{E}_{i, Z} \right\}
\]
where $\mathfrak{B}_{i, Z} = (x_{i,Z,0}, \dots, x_{i,Z,n+1})$ are the histogram bin edges (in \qty{}{\MeV\per\atomicmassunit}), and $\mathfrak{E}_{i,Z} = (y_{i,Z,0}, \dots, y_{i,Z,n})$ are the corresponding normalized bin contents, in units of \si{\count \per \primary \per \centi\meter\squared \per \MeV \atomicmassunit}. Thus, $\widetilde{E}_{i,Z}$ represents the normalized differential spectrum at fixed configuration $i$ and charge number $Z$. The same  binning is used for all spectra (target and modulator simulations), which facilitates direct comparisons. 
For each charge number $Z$, the spectra from all six configurations are collected in matrix form: 
\[
\mathbb{E}_{Z} =
\begin{pmatrix}
\mathfrak{E}_{0,Z} \\
\mathfrak{E}_{1,Z} \\
\vdots \\
\mathfrak{E}_{5,Z}
\end{pmatrix}
\in \mathbb{R}^{6 \times n},
\]
while the corresponding target spectrum is written as: 
\[
T_{Z} = \begin{pmatrix}
\hat{y}_{Z,0} & \hat{y}_{Z,1} & \dots & \hat{y}_{Z,n}
\end{pmatrix} \in \mathbb{R}^{1 \times n}.
\]

The optimization problem is then: 
\[
\arg \min_{\omega} \; \left\{ f(\mathbb{E}_{Z}, \omega, T_{Z}), \; \forall Z \right\},
\]
where $\omega \in \mathbb{R}^{1 \times 6}$ is the vector of weights assigned to each configuration, and $f$ is the cost function.
To preserve physical interpretability and allow a more granular control, the cost function is evaluated in two stages. First, for each $Z$:
\[
\phi_{Z}(\omega) = \frac{\big(\omega \mathbb{E}_{Z} - T_{Z}\big)^2}{T_{Z}} \in \mathbb{R}^{1 \times n},
\]
which quantifies the normalized bin-wise deviation between the weighted spectra and the target. Each $\phi_{Z}$ explicitly depends on the weight vector $\omega$, expressed in units of \si{\primary \per \second}. This step ensures that the contributions of each configuration are optimized simultaneously across all energy bins for each ion.
Second, deviations are summed over all charge numbers with optional user-defined ``importance'' factors $w_Z$, which weight each ion species according to its relative contribution to the target absorbed dose:

\[
\mathcal{E}(\omega) = \sum_{Z=1}^{26} w_{Z}\,\phi_{Z}(\omega).
\]

Similarly, the total target spectrum is defined as:

\[
\mathcal{T} = \sum_{Z=1}^{26} T_{Z},
\]
and its statistical uncertainty across species can be estimated as:

\[
\tilde{\sigma}^2 = \langle T_{Z}^2 \rangle_Z - \langle T_{Z} \rangle_Z^2.
\]

A scalar objective function is finally obtained by normalizing the summed deviations by the target standard deviation:

\[
\xi(\omega) = \sum_{k=0}^{n} \frac{\mathcal{E}(\omega) - \mathcal{T}}{\tilde{\sigma}} \, .
\]

This two-stage optimization structure guarantees that both individual kinetic-energy spectra and the total expected flux across all species are reproduced.\\
The minimization is carried out in \texttt{Python} using the \texttt{SciPy} ``trust-constr'' \cite{Conn2000} solver. Some of the key advantages of this algorithm include the ability to constrain the weights, set tolerances, and define convergence parameters. These parameters were fine-tuned to keep the minimization within a reasonable time while maintaining a good convergence. On a standard desktop, the full optimization converges in 3.5 seconds, after 350 iterations and 2149 function evaluations. 
This performance is order of magnitudes faster than the underlying Monte Carlo simulations. As a result, computation times of a few seconds up to a few tens of seconds are entirely acceptable for this application, and do not limit tuning or operational reproducibility of the GCR simulator.

\section{Kinetic energy spectra}
\label{app:kinetic_energy_spectra}

This appendix presents the kinetic energy spectra for all particle species, from $Z=0$ (neutrons) to $Z=26$ (iron). 
Each figure compares the Target spectrum (light blue) with the GCR simulator spectrum (dark blue) obtained after the optimization process, showing how closely the simulator reproduces the reference distributions for each ion species in the kinetic energy range between \qty{10}{\MeV \per \atomicmassunit} and \qty{2000}{\MeV \per \atomicmassunit}.\\ 
As noted in \autoref{ssec:GCRSimulator_Optimization_process}, neutrons were \textit{not directly optimized}; their kinetic energy spectrum results from the weights found for charged species (protons to iron) and is reported here for completeness.\\
Neutrons are abundantly produced as secondary particles and are relevant for experiment design at the GSI's GCR simulator. However, they were not included in the optimization process, and total differential fluxes, LET spectra, or dose calculations were performed only for charged particles.

\foreach \Z in {0,...,26}{%
    \begin{figure}[H]
    \centering
    \includegraphics[width=0.7\columnwidth]{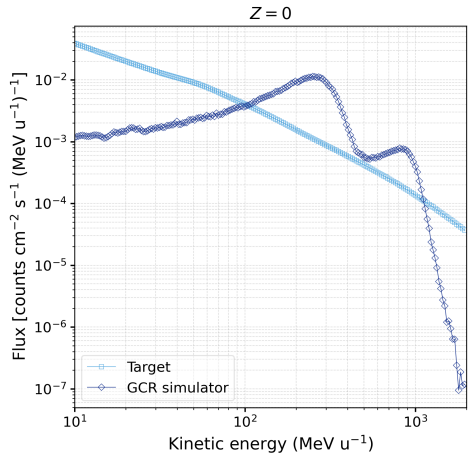}
    \caption{Kinetic energy spectra for $Z=\Z$.}
    \end{figure}
}

\end{document}